\documentclass[aps,amsmath,amssymb,twocolumn]{revtex4}
\usepackage{graphics, setspace}
\usepackage[letterpaper, dvips,width=7.5in,height=8.5in,includemp=false]{geometry}

\newcommand{\be}{\begin{equation}}
\newcommand{\ee}{\end{equation}}

\begin{document}

\title{Comment on the Walliser-Weigel approach to exotic baryons in chiral soliton models}

\author{Thomas D. Cohen}
\email{cohen@physics.umd.edu}

\affiliation{Department of Physics, University of Maryland, College
Park, MD 20742-4111}

\begin{abstract}
This comment discusses a recent paper by Walliser and Weigel on the
quantization of chiral soliton models in the context of exotic
baryons.  Claims made in that work are misleading due to unfortunate
nomenclature. Moreover, attempts in that paper to go beyond the
leading order calculations of the phase shifts are {\it ad hoc} and
never justified.  This comment also addresses a technical issue in
that paper: the identification of the excitation energy of the
pentaquark obtained via conventional rigid rotor quantization with a
frequency obtained in the context of small amplitude fluctuations.
The identification is erroneous: the small amplitude fluctuation
result is based on a first-order perturbation computation of the
frequency around a zero mode solution at a frequency far from zero
and well away from the perturbative regime.
\end{abstract}

\maketitle

The comment seeks to clarify the recent paper by Walliser and
Weigel\cite{ww} on the quantization of chiral soliton models in the
context of exotic baryons.   A principal purpose of this comment is
to demystify several statements in ref. \cite{ww}  which are
extremely misleading.

Ref.~\cite{ww} uses the phrases ``rigid rotor quantization'' and
``collective''  differently than is used in much of the literature.
This may create the false impression that basic fabric of the
approach is in conflict with earlier analyses
\cite{Cohen:2003yi,Cohen:2003mc,Itzhaki:2003nr,Pobylitsa:2003ju,Cherman:2004qx,latest}.
These earlier analyses showed that the rigid-rotor approach as
employed by Praszalowicz\cite{Praszalowicz} and by Diakonov, Petrov
and Polyakov\cite{Diakonov:1997mm} is not justified at large $N_c$
for baryons with exotic quantum numbers and that the correct way at
large $N_c$ to compute the relevant physical observable, namely, the
phase shift, is via small amplitude fluctuations ({\it i.e.}, the
method of Callan and Klebanov\cite{CK}. In fact, the explicit
calculations by Walliser and Weigel show precisely that the Callan
and Klebanov approach gives the exact phase shifts at large $N_c$
and that the rigid-rotor approach used by Praszalowicz and by
Diakonov, Petrov and Polyakov cannot be used validly without
substantial modification. Walliser and Weigel refer to this
modification as the rotation-vibration approach (RVA).  It is shown
in sec. V of ref.~\cite{ww} that at large $N_c$ the RVA is
indistinguishable from the Callan-Klebanov approach. Thus, the
substance of ref.~\cite{ww} is in accord with the analyses of
refs.~\cite{Cohen:2003yi,Cohen:2003mc,Itzhaki:2003nr,Pobylitsa:2003ju,Cherman:2004qx,latest}.

However, despite this fundamental agreement of
refs.~\cite{Cohen:2003yi,Cohen:2003mc,Itzhaki:2003nr,Pobylitsa:2003ju,Cherman:2004qx,latest}
with the approach of Walliser and Weigel for the $S$-matrirx at
large $N_c$, multiple statements in ref.~\cite{ww} appear to be in
conflict with
refs.~\cite{Cohen:2003yi,Cohen:2003mc,Itzhaki:2003nr,Pobylitsa:2003ju,Cherman:2004qx,latest}.
As an example, consider the claim in the abstract of
ref.~\cite{ww} that ``{\it We thoroughly compare the bound state}
[that is the Callan-Klebanov approach] {\it and rigid rotor
approaches to three-flavored chiral solitons. We establish that
these two approaches yield identical results for the baryon
spectrum and the kaon-nucleon S-matrix in the limit that the
number of colors tends to infinity.}'' This appears to be in flat
contradiction that the rigid rotor approach gives incorrect
results for exotic states and the correct approach is to compute
phase shifts according to the Callan-Klebanov approach as was
originally done in ref.~\cite{Itzhaki:2003nr}.

A central reason for this apparent discrepancy is linguistic. When
refs.~\cite{Cohen:2003yi,Cohen:2003mc,Itzhaki:2003nr,Pobylitsa:2003ju,Cherman:2004qx,latest}
refer to the ``rigid rotor approach'' they mean the approach used by
Praszalowicz\cite{Praszalowicz} and by Diakonov, Petrov and
Polyakov\cite{Diakonov:1997mm} in the early predictions of
pentaquarks from chiral soliton models. That is, they mean using the
collective Hamiltonian introduced by Guadagnini \cite{Guadagnini} to
directly calculate properties of discrete states which are then
equated with the pentaquark. In the language of ref.~\cite{ww} this
corresponds to simply using the Lagrangian of Eq.~(4.1), converting
it to a collective Hamiltonian, quantizing the Hamiltonian and using
the discrete states so obtained to calculate directly the physical
of baryons.   The central purpose of
refs.~\cite{Cohen:2003yi,Cohen:2003mc,Pobylitsa:2003ju,Cherman:2004qx,latest}
was to show that this procedure while legitimate for the two flavor
Skyrmion and for non-exotic states in the three flavor models, is
inadequate to compute the physical properties of exotic baryons in
the large $N_c$ limit of the theory. It is clear that the authors of
ref.~\cite{ww} agree that it is not. They write in sec. I B that
``{\it ...it is necessary to include small amplitude fluctuations in
the RRA.  We call this approach the rotation-vibration approach
(RVA).}''  The authors then go on to show, correctly, that the
S-matrix computed in the RVA is identical to that computed
Callan-Klebanov approach at large $N_c$. Thus, when claims are made
in ref. \cite{ww} that the ``rigid-rotor approach'' agrees with the
Callan-Klebanov approach at large $N_c$, it is often referring to
the RVA and not to the unadulterated rigid rotor approach used by
Praszalowicz\cite{Praszalowicz} and by Diakonov, Petrov and
Polyakov\cite{Diakonov:1997mm}.  Thus, despite the apparent
conflict, there is no underlying disagreement between the results of
these previous analyses and the large $N_c$ results of Walliser and
Weigel for the physical observables.

Of course, as a logical matter, definitions when used in a
self-consistent manner cannot be incorrect. In this sense, there is
nothing wrong with ref.~\cite{ww} use of ``rigid rotor approach'' to refer
to a calculation including both vibrations and rotations. However,
it would have been wise to avoid such a usage: it is seriously
misleading about previous work.  It is also a particularly perverse
locution in that the vibrations, the non-rigid degrees of freedom,
play an essential role.

The authors of ref.~\cite{ww} use of the word ``collective''  can
also be seriously misleading.  Part of the origin of this
potential confusion stems from the fact that ``collective'' has
many usages in the literature.  One common meaning in many-body
physics is to refer to dynamics in which many particles are moving
with significant coherence: {\it eg.}, the giant dipole resonance
in low energy nuclear physics.  A ``collective coordinate'' in
such a context refers to a choice of coordinate which couples
strongly to this coherent motion and weakly to the single particle
motion.  The definition of such a degree of freedom is somewhat
arbitrary in such circumstances but a judicious choice can
simplify one's description. More generally, one can introduce
``collective coordinates'' in a similar sense for problems in
which some class of motion which is characteristically more
coherent than typical motion.

There is also a second, more technical, usage of ``collective''
which is commonly used in semiclassical treatments of soliton
models\cite{soliton}.  In this context, ``collective'' refers to
degrees of freedom which are formally adiabatic in the sense that
they become arbitrarily slow as the weak coupling ({\it i.e.},
semiclassical) limit is approached.  Such modes are associated with
the zero modes of the linearized classical equations of motion
around static solitons. They have the property that they completely
decouple from all non-collective degrees of freedom in the weak
coupling limit. The large $N_c$ limit is precisely such a weak
coupling expansion. Thus, the formulation of a systematic $1/N_c$
expansion for chiral soliton models depends on correctly identifying
the ``collective modes'' in this second more restricted sense.
References~\cite{Cohen:2003mc,Cherman:2004qx,latest} carefully
distinguish between static and dynamic zero modes and show that not
all static zero modes have associated dynamic zero modes.  This is
done precisely to show there are fewer collective degrees of freedom
in this second more technical sense than are obtained by counting
the number of flat directions assuming that there is one pair of
collective modes for each.  This, in turn is extremely useful in
developing a simple systematic treatment of the theory at large
$N_c$ and zero SU(3) flavor breaking.

When Walliser and Weigel write, ``{\it Generally we may always
introduce collective coordinates to investigate specific
excitations.  This is irrespective of whether the corresponding
excitation energies are suppressed in the large $N_c$ limit or not.
This is in contrast to the claims of ref. [33]}  [refs.
\cite{Cohen:2003yi,Cohen:2003mc} of this comment] {\it that a scale
separation is necessary to validate the collective quantization
coordinate quantization. Eventually the coupling terms and
constraints ensure correct results.}'' and ``{\it ...collective
coordinates are {\em not} necessarily linked to zero modes of a
system and any distinction between dynamical and static zero–modes
[34]} [ref. \cite{latest} of this comment] {\it is obsolete.}'',
they {\it must} be referring to some variant of the first usage of
``collective'' and not the second more technical sense.  The
statement is manifestly wrong if used in the second sense. Moreover,
in this context the phrase ``collective coordinate quantization''
used above clearly refers to the RVA (which as noted above is at
large $N_c$ identical to the Callan-Klebanov approach) and not the
rigid rotor approach of Guadagnini \cite{Guadagnini}. This is
unfortunate since for ``collective quantization''  may call to mind
Guadagnini's rigid rotor approach and thus create an impression at
odds with the facts.

Before turning to the substance of ref.~\cite{ww}, a final quibble
about language.   Reference \cite{ww} uses the unfortunate moniker
``bound state approach'' to describe the Callan-Klebanov method of
small amplitude vibrations. While Callan and Klebanov did refer to
this method as the ``bound state approach'' \cite{CK}, that was in
the context of non-exotic states which were, in fact, bound. For
exotic states---which are unbound resonances---describing the
method as the ``bound state approach'' is perverse.  For this
reason this method will be referred to as the ``Callan-Klebanov''
approach throughout this comment.

Apart from these  linguistic issues, there are substantial
problems in some of the calculations in ref.~\cite{ww}: important
conceptual difficulties arise in attempts to compute quantities
associated with exotic baryons beyond the computation of the phase
shifts at leading order in $1/N_c$---the regime in which it agrees
with
refs.~\cite{Cohen:2003yi,Cohen:2003mc,Itzhaki:2003nr,Pobylitsa:2003ju,Cherman:2004qx,latest}.
Two principal issues arise in this regard: i) the separation of
the phase shift into a ``background'' and ``resonant'' part, and
ii) the calculation of the width of the exotic baryon for $N_c=3$.

Consider issue i):  the paper relies on a separation of the phase
shift into a ``background'' and ``resonant'' part to conclude that
there is a resonance in the exotic pentaquark channel. However,
this separation is unnecessary, pointless, arbitrary and poorly
motivated.  It is unnecessary since the phase shifts (which are
directly computed in the Callan-Klebanov approach) contain {\it
all} of the physics about the scattering available at leading
order. It is pointless since there is no possible experiment which
measures the ``resonant" piece separately. It is arbitrary, since
{\it any} separation of a full amplitude into a background and
resonant contribution is ultimately model dependent and hence
arbitrary. Finally, it is poorly motivated: the ``background''
depends on a misidentified collective degree of freedom.  It
associates the background with those fluctuations orthogonal to
the ``collective'' fluctuation of freedom. However, as will be
discussed in detail below, this putative ``collective''
fluctuation plays no special role either physically or
mathematically. It should be stressed, however, that regardless of
whether one believes that there is some mathematical or physical
significance to the ``collective'' mode identified for exotic
baryons, the separation of the phase shift into ``background'' and
``resonant'' contributions remains unphysical.

Next focus on the motivation for separating the phase shift into a
``background'' and ``resonant'' part.  The background is taken to be
due to fluctuations constrained to be orthogonal to the
``collective'' mode identified by the authors.  This ``collective''
mode is the zero mode associated with non-exotic oscillations,
$z(r)$. As noted above, there are two commonly used meanings of
``collective'', but the mode taken as ``collective'' here appears to
satisfy neither definition when used to describe the exotic degree
of freedom. It is neither associated with a zero mode of the exotic
channel as in the technical definition used in
refs.~\cite{Cohen:2003mc,Cherman:2004qx,latest} nor does it
correspond approximately to a particularly coherent class of motion
which dominates the behavior in some region. Moreover, it is not
special mathematically except at zero frequency (where it satisfies
the small amplitude equation of motion and is associated with the
non-exotic fluctuation); at finite frequencies in the neighborhood
of the putative resonance it does {\it not} satisfy the small
amplitude equation of motion.  This issue will be discussed in some
detail at the end of this comment. Thus, there is no apparent
special quality about this mode, neither physically nor
mathematically.  It seems to be a completely arbitrary choice and
there is no apparent reason to identify the motion orthogonal to it
as corresponding to ``background''.

Next consider issue ii): the claims the pentaquark width is
computed at $N_c=3$. However, this claim is unjustified on very
basic grounds.  In the first place, it is based on the
``resonant'' phase shift and as noted above the separation into
``resonant'' and ``background'' is unjustified.  However, even
were that not the case, the calculation at $N_c=3$ would not be
legitimate. Note that entire formalism for treating the chiral
soliton model is based on a $1/N_c$ expansion and has only been
computed consistently at the leading nontrivial order.  For
example, the form of the profile function $F(r)$ was computed
using classical equations of motion which are valid at large $N_c$
but which have subleading corrections. At $N_c=3$ these induce
corrections which are not included in this paper. Similarly, the
computation of kaon-nucleon $S$ matrix does not include dynamical
effects in which the kaon-nucleon scattering goes into
kaon-pion-nucleon states. Such effects alter the kaon-nucleon
scattering amplitude when functioning as intermediate states and
for energies above the pion production amplitude yields inelastic
contributions. One can justify dropping this dynamics in the large
$N_c$ limit but not at $N_c=3$.  Thus, the authors make one set of
$1/N_c$ corrections---those associated with the  ``collective''
mode---to all orders to get an $N_c=3$ result while simultaneously
neglecting even the first-order correction to others. {\it A
priori} there is no reason why such a result should be any more
reliable at $N_c=3$ than the leading order expression. Certainly,
the $1/N_c$ expansion does not justify such a procedure and the
authors give no other argument.

Thus, both the procedure to separate the ``background'' and
``resonant'' contributions and the procedure to compute the
resonant width at $N_c=3$ are {\it ad hoc}.  Neither has been
justified in a systematic $1/N_c$ expansion nor from any other
systematic framework. Any conclusion based on these must viewed
with skepticism.

The final issue addressed in this comment is an error in
ref.~\cite{ww} which can lead to significant confusion.  The
context is the small amplitude equation for the kaon vibrational
modes around a Skyrmion imbedded in the $u$-$d$ subspace in Eq.
(3.5) of ref.~\cite{ww}:
\begin{equation}
h^2 \eta (r) + \omega \left  [ 2 \lambda (r) - \omega M_{K} (r)
\right ] \eta(r) \; ,
\label{vib}
\end{equation}
where $h^2(r)$  is a differential operators, $\lambda (r)$ and
$M_K(r)$ are functions; $\lambda$ arises from the Witten-Wess-Zumino
term.  There is a zero frequency solution to this equation in the
limit of zero SU(3) flavor breaking:
\begin{equation}
z(r) = \frac{ \sqrt{4 \pi}  f_\pi \sin \left ( F(r)/2 \right
)}{\sqrt{\Theta_K} }
\end{equation}
where $F(r)$ is the Skyrmion profile function and $\Theta_K$ is
the normalization constant computable from $F(r)$. Now suppose one
wants to include SU(3) violations into this formalism: the SU(3)
violating term is in $h^2(r)$ and equals $m_K^2 -m_\pi^2$. For
small SU(3) violations one can use first-order perturbation theory
to compute the shift of the frequency of the zero mode away from
zero. This is done by simply taking Eq.~(\ref{vib}) replacing
$\eta(r)$ by $z(r)$, multiplying on the right by $z(r)$ and
integrating.  One obtains
\begin{eqnarray}
\omega^2 & = & \frac{3 \Gamma}{8 \Theta_K} + \omega_0 \omega \; ,
\label{pert}\\
{\rm with} \; \; \omega_0 & \equiv & \int \, d r \, r^2 z^2(r) 2 \lambda(r) = \frac{N_c}{4 \Theta_K} \; , \\
{\rm and } \; \; \Gamma & \equiv & \frac{8 \Theta_K (m_K^2 - m_\pi^2)}{3} \int \, d r \, r^2 z^2(r) \; ,\nonumber \\
\end{eqnarray} which corresponds to Eqs. (3.7)-(3.9) of
ref.~\cite{ww}.  Since Eq. (\ref{pert}) was obtained from the zero
mode via first-order perturbation theory in SU(3) breaking it is
only valid at linear order in $\Gamma$ and has the unique solution
at this order:
\begin{equation}
\omega \equiv \omega_\Lambda = \frac{3 \Gamma}{8 \Theta_K
\omega_0} + {\cal O} (\Gamma^2). \label{firsto}\end{equation}
The
frequency is denoted by $\omega_\Lambda$ as it corresponds to the
excitation energy of the $\Lambda$ above the nucleon.

While ref.~\cite{ww} has Eq. (\ref{pert}) (it is Eq.~(3.9) of that
reference), it neglects to mention that the equation obtained from
the zero mode via first-order perturbation theory in SU(3) breaking
and thus is only valid near zero and only to first order in
$\Gamma$. This is unfortunate since ref.~\cite{ww} uses Eq.~(3.9) to
all orders in $\Gamma$ to obtain---incorrectly---a general
expression for $\omega_\Lambda$.  Even more troubling,
ref.~\cite{ww} identifies the second solution of the quadratic in
Eq. (\ref{pert}) with the $\theta^+$ pentaquark:
\begin{equation}
\omega_\theta=\left ( \sqrt{\omega_0^2 + \frac{3 \Gamma}{2
\Theta_K}} + \omega_0 \right )/{2} \; ,\label{allord}\end{equation}
which is Eq. (3.11) of ref.~\cite{ww}.  However, the calculation of
$\omega_\theta$ is totally without justification. It is based on
perturbation theory around the zero mode and is only valid near
$\omega=0$. However, the solution is parametrically far from
$\omega=0$ (it is of order $N_c^0$ regardless of the value $\Gamma$)
and hence outside the regime of validity of the perturbative
treatment. One can see that the claimed result is wrong simply by
returning to the SU(3) limit in which case $\omega_\theta=\omega_0$.
If one focuses on the original small amplitude
equation---Eq.~(\ref{vib}) above---one sees explicitly that $z(r)$
is {\it not} an eigenfunction corresponding to an eigenfrequency of
$\omega_0$. In summary, the quoted expression for $\theta_\theta$ is
unjustified in the context of the Callan-Klebanov approach.
Moreover,  it is clear in this analysis that $z(r)$, while a
collective degree of freedom associated with a zero mode for the
non-collective state,  plays no special collective role for the
exotic channel---it merely represents a mode for which was chosen
{\it arbitrarily} as the mode in which the expectation value of
Eq.~(\ref{vib}) was taken.

The erroneous calculation of $\omega_\theta$ leads to an important
misstatement  in ref.~\cite{ww}.  In the introduction, it is stated
that ``{\it In sections III and IV we compare the two approaches}
[the rigid rotor approach and the Callan-Klebanov approach] {\it and
show how they yield identical spectra at large $N_c$...}''.  Note in
the context of sec. IV of ref.~\cite{ww} the rigid rotor approach
refers to the original approach of Guadagnini\cite{Guadagnini} and
not the RVA.  Thus, if this statement were correct, the analysis of
refs.~\cite{Cohen:2003yi,Cohen:2003mc,Itzhaki:2003nr,Pobylitsa:2003ju,Cherman:2004qx,latest}
would be wrong.  However, the evidence presented for this in the
context of exotic states at zero SU(3) violation is given in
Eq.~(14) where it is shown that at large $N_c$ the excitation of the
pentaquark is identical with $\omega_\theta$ as calculated above.
From this the authors write, ``{\it Thus we conclude the BSA} [i.e.,
Callan-Klebanov approach] {\it and the RRA are consistent when
flavor symmetry breaking is omitted.}''  This conclusion is
unjustified. It is based on an invalid calculation of
$\omega_\theta$ in the Callan-Klebanov approach.  Indeed, based on
the calculation, the authors ought to have concluded that the two
methods are, in fact, distinct.

In summary, the approach of Walliser and Weigel in ref.~\cite{ww},
where justified by the $1/N_c$ expansion, agrees with the
Callan-Klebanov approach and is consistent with the analyses of
ref.~\cite{Cohen:2003yi,Cohen:2003mc,Itzhaki:2003nr,Pobylitsa:2003ju,Cherman:2004qx,latest}.
However, this basic fact is generally obscured by a very unfortunate
use of language in refs.~\cite{ww}.  It is obscured further by the
erroneous of computation $\omega_\theta$ which is then used to argue
spuriously that the rigid rotor approach and the Callan-Klebanov
approach are equivalent. Finally, the principal new elements of this
approach, the separation of the phase shifts into a ``background''
and ``resonant'' part, and the scheme to compute the phase shifts at
$N_c=3$ are not justified.

Support of the U. S. Department of Energy under grant no.
DE-FG02-93ER-40762 is gratefully acknowledged.

\bibliographystyle{amsplain}

\end{document}